\documentclass[preprint, 12pt]{aastex}
\usepackage{amsmath,amssymb}

\newcommand{\chandra}{\textit{Chandra}}
\newcommand{\xmm}{\textit{XMM}}
\newcommand{\zpup}{$\zeta$ Puppis}

\newcommand{\taustar}{\tau_*}
\newcommand{\Rstar}{R_*}
\newcommand{\hetgs}{HETGS}
\newcommand{\meg}{MEG}
\newcommand{\etal}{et al.}
\newcommand{\vinf}{v_\infty}
\newcommand{\Ro}{R_\mathrm{o}}

\newcommand{\msunyr}{\ensuremath{\mathrm{M}_\sun~\mathrm{yr}^{-1}}}
\newcommand{\mdot}{\dot{M}}

\shorttitle{\zpup\ Line Profile Modeling}
\shortauthors{Kramer, Cohen, \& Owocki}

\begin{document}

\title{X-ray Emission Line Profile Modeling of O Stars:\\ Fitting a
Spherically-Symmetric Analytic Wind-Shock Model to the {\it Chandra}
Spectrum of $\zeta$ Puppis}
\author{Roban H. Kramer\altaffilmark{1, 2}, David
H. Cohen\altaffilmark{1}, Stanley P. Owocki\altaffilmark{3}}
\email{roban@sccs.swarthmore.edu; cohen@astro.swarthmore.edu;
owocki@bartol.udel.edu}
\altaffiltext{1}{Swarthmore College Department of Physics and
 Astronomy, 500 College Ave., Swarthmore PA 19081}
\altaffiltext{2}{Prism Computational Sciences, 455 Science Dr.,
Madison WI 53711}
\altaffiltext{3}{Bartol Research Institute, University of Delaware,
 217 Sharp Laboratory, Newark DE 19716}

\begin{abstract}
X-ray emission line profiles provide the most direct insight into the
dynamics and spatial distribution of the hot, X-ray-emitting plasma
above the surfaces of OB stars. The O supergiant \zpup\ shows broad,
blueshifted, and asymmetric line profiles, generally consistent with
the wind-shock picture of OB star X-ray production. We model the
profiles of eight lines in the \chandra\ \hetgs\ spectrum of this
prototypical hot star. The fitted lines indicate that the plasma is
distributed throughout the wind starting close to the photosphere,
that there is significantly less attenuation of the X-rays by the
overlying wind than is generally supposed, and that there is not a
strong trend in wind absorption with wavelength.
\end{abstract}


\section{Introduction}

The nature of the copious soft X-ray emission from hot stars has been
a longstanding controversy since its discovery in the late 1970s
\citep{h79,co79}. Solar-type coronal emission was first assumed
\citep{co79,w84}, but wind-shock models of various types gained
currency throughout the following decade
\citep{w84,ocr88,mc89,cw91,hillier93,cohen96,feldmeier97,fpp97}. More
recently, hybrid magnetic wind models have been proposed for some hot
stars \citep{g97,bm97,uo02}.

Until the launch of \chandra\ and \xmm, no observational diagnostics
were available that could provide a direct discriminant between the
coronal and wind-shock paradigms. But due to their superior spectral
resolution, both of these new telescopes allow for the separation of
individual emission lines and the resolution of Doppler-broadened line
profiles. The spectral resolution of \chandra's grating spectrometers
exceeds $\frac{\lambda}{\Delta \lambda} \sim 1000$ (for the FWHM)
which corresponds to a velocity of $300$ km s$^{-1}$, and that of the
\xmm\ RGS is almost as great. This compares favorably to the terminal
velocities of the radiation-driven winds of O stars, which approach
$\vinf = 3000$ km s$^{-1}$, implying $\sim 20$ resolution elements
for a velocity range of $2\vinf$.

At the most basic level, the X-ray emission lines from hot stars will
either be narrow and therefore roughly consistent with coronal
emission or broad and roughly consistent with wind-shock
emission. This is because in the wind-shock model the high velocities
of X-ray-emitting plasma embedded in the wind would Doppler-shift the
emission across a range of wavelengths. Initial papers reporting on
\chandra\ and \xmm\ observations of various O and B stars
\citep{s01,k01,wc01,cass01,m03,cohen03} discussed line widths, which
vary from large for early O stars to small for B stars.  Some of these
initial studies noted that the emission lines can be blueshifted and
the profiles somewhat asymmetric, but none of these studies discussed
or modeled the shapes of the resolved emission lines.

In this paper we fit a specific model of X-ray emission line profiles
in an expanding, emitting, and absorbing wind to a \chandra\
\hetgs/\meg\ spectrum of \zpup. The model we fit is empirical and
flexible, with only three free parameters. The model assumes a
two-component fluid, having as its major constituent the cold,
X-ray-absorbing plasma that gives rise to the characteristic UV
absorption lines observed in hot star winds, and as its minor
constituent the hot, X-ray-emitting plasma. This empirical model is
not tied to any one specific physical model of X-ray production, and
is general enough to fit data representative of any of the major
models, so long as they are spherically symmetric. To the extent that
wind-shock models are found to be consistent with the observed line
profiles, our model parameters can be used to constrain the physical
properties of the shock-heated plasma. This ultimately can be used to
constrain the values of the physical parameters of the appropriate
wind-shock model.

In section 2 we discuss the physical effects leading to non-trivial
line shapes and describe the specific empirical model we use to
perform the fits. In section 3 we describe the \chandra\ \zpup\
dataset and how we perform the line fitting and parameter
estimation. And in section 4 we discuss the derived model parameters
in the context of the various physical models that have been proposed
to explain hot-star X-ray emission, as well as in the context of other
X-ray diagnostics that have recently been applied to the data from
this star.

\section{Theoretical Considerations}

In the context of the fast, radiation-driven winds of OB stars, a
source of X-ray emission embedded in the wind will lead to Doppler
broadened profiles, but only the X-ray emitting plasma traveling at
the wind terminal velocity directly toward or away from the observer
will lead to maximal blueshifts and redshifts. The amount of emission
at each intermediate wavelength, and thus the shape and characteristic
width of the line, depends on the spatial and velocity distribution of
the hot plasma. As described by \citet{m91} in the case of a
shell of X-ray emitting plasma in a hot star wind, the continuum
absorption of X-rays by the cool component of the wind will cause the
resulting emission lines to be attenuated on the red side and be
relatively unaffected on the blue side of the line profile. The
apparent peak of the emission line thus shifts to the blue side of
line center, and the line is asymmetric, with a shallower red wing and
a steeper blue wing.

This basic idea of Doppler broadened emission from a hot wind
component and continuum absorption by the cool wind component leading
to a broadened, shifted, and asymmetric line was extended from a shell
to a spherically symmetric wind by \citet{i01}. He showed that model
line profiles could be generated analytically for a constant-velocity
wind. \citet{oc01} extended this concept further, to an accelerating
wind, with a model having four free parameters. Two describe the
spatial distribution of the X-ray emitting plasma -- $\Ro$, the
minimum radius of X-ray emission and $q$, the radial power-law index
of the emissivity. There is assumed to be no emission below $r =
\Ro$. Above $r = \Ro$, volume emissivity is assumed to scale like the
density of the wind squared $\rho^2$ (since collisional processes and
recombination dominate the ionization/excitation kinematics), with an
extra factor $r^{-q}$ allowing for spatial variation of shock
temperatures, cooling structures, and density and filling factor of
the shocked material. The parameter $\beta$ controls the velocity of
the wind, which is assumed to follow a ``beta-velocity law'':
\begin{equation}\label{v_eq}
v(r) = \vinf(1-\Rstar/r)^\beta \;\mathrm{.}
\end{equation}
Both components of the wind follow the same velocity law in all cases
discussed here, but in principle they could be allowed to differ. The
fourth parameter, $\taustar$, characterizes the amount of absorption
in the wind
\begin{equation}\label{taustar_eq}
\taustar \equiv \frac{\kappa \mdot}{4 \pi \vinf \Rstar} \text{ ,}
\end{equation}
where $\kappa$ is the line opacity or mass absorption coefficient
(cm$^2$~s$^{-1}$). This relates to the commonly-quoted radius of
optical depth unity $R_1$ by the equation (for $\beta = 1$)
\begin{equation}\label{r1_eq}
\frac{R_1}{\Rstar} = \frac{1}{1-\exp{(-1/\taustar)}} \approx \taustar
+ 0.5 \text{ for $\taustar > 0.5$.}
\end{equation}
The error in the above approximation is less than 10\% for $\taustar
> 0.67$ (or $R_1/\Rstar > 1.3$).

The model put forward by \citet{oc01} is a phenomenological one. It
describes the physical properties of the hot and cool components of
the wind, but does not describe the physics underlying the generation
of the hot plasma. It is therefore quite flexible, capable of
describing a thin shell of X-ray emitting plasma, including a
coronal-type zone near the photosphere, as well as wind shocks
distributed spatially throughout the wind, with the shock distribution
and wind velocity varying with radius. This model is thus capable of
constraining properties of both the X-ray-emitting and the
X-ray-absorbing wind components. When applied to an ensemble of lines
it has the potential to constrain these properties as a function
of both temperature and wavelength. Assuming turbulent and thermal
broadening are negligible and combining the other model assumptions,
the line profile as a function of scaled wavelength $x \equiv
(\lambda/\lambda_0 - 1)(c/\vinf)$ is given by
\begin{equation}\label{L_eq}
L_\mathrm{x} \propto \int_{r=r_x}^{\infty} \frac{r^{-(q+2)}}{\left( 1
- \frac{\Rstar}{r} \right)^{3\beta}} \exp\left[-\tau \left(\mu_x,\; r \right) \right] dr \text{ ,}
\end{equation}
where $r_x \equiv \max\left[\Ro,\Rstar/(1-|x|^{(1/\beta)})\right]$,
$\mu_x \equiv x/(1-\Rstar/r)^\beta$ and $\tau\left(\mu, r\right)$
(which is proportional to $\taustar$) is the optical depth along the
observer's line of sight at direction cosine $\mu$ and radial
coordinate $r$. The constant of proportionality (which is, itself,
proportional to the emission measure) will not be determined in this
work. Equation (\ref{L_eq}) must be solved numerically, except in the
case of $\beta = 0$ (constant velocity wind). And even so, we only
obtain solutions for integer values of $\beta$.

\begin{figure}
\includegraphics[scale=0.35]{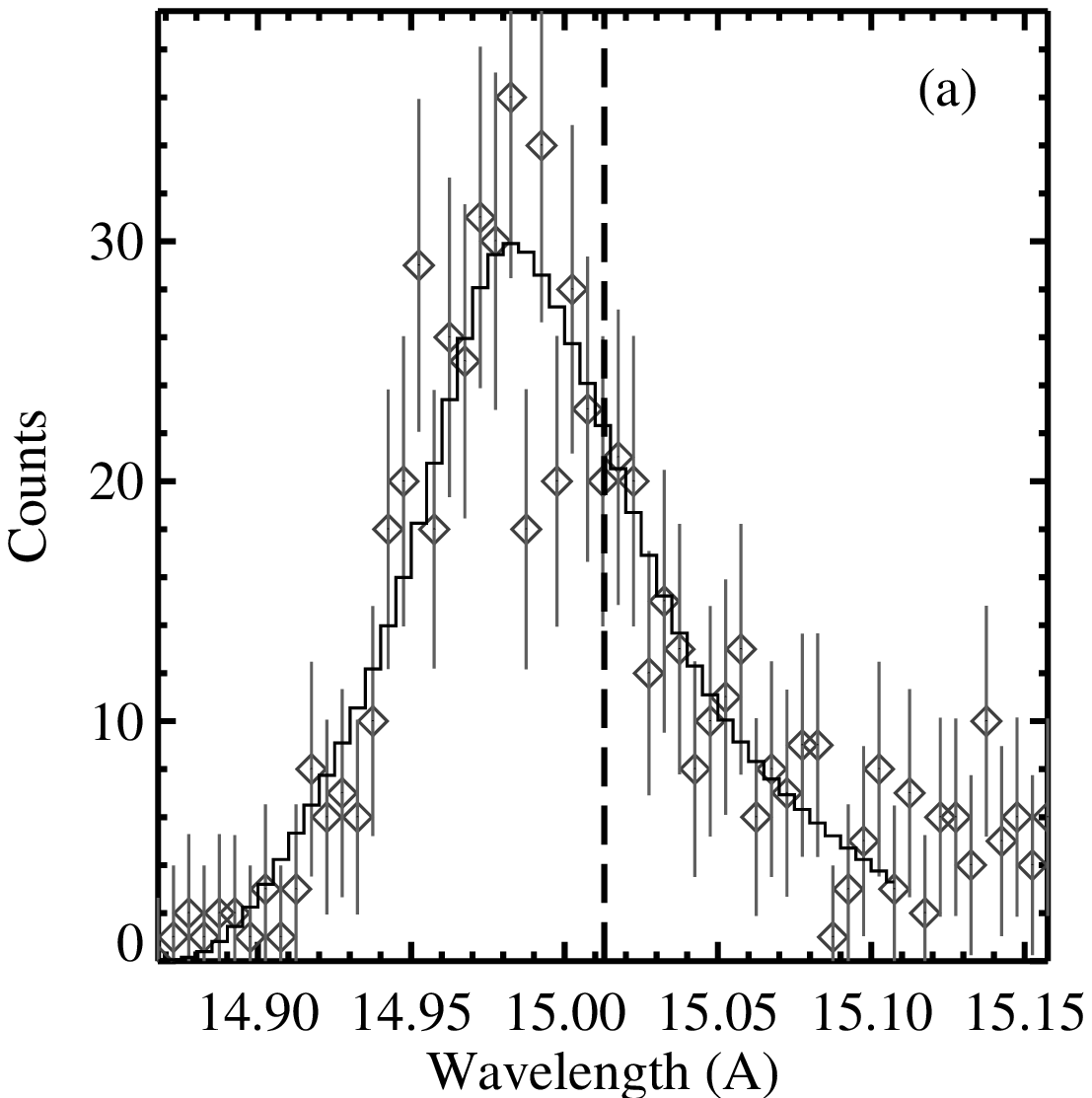}
\includegraphics[scale=0.3]{f1b.eps}
\includegraphics[scale=0.35]{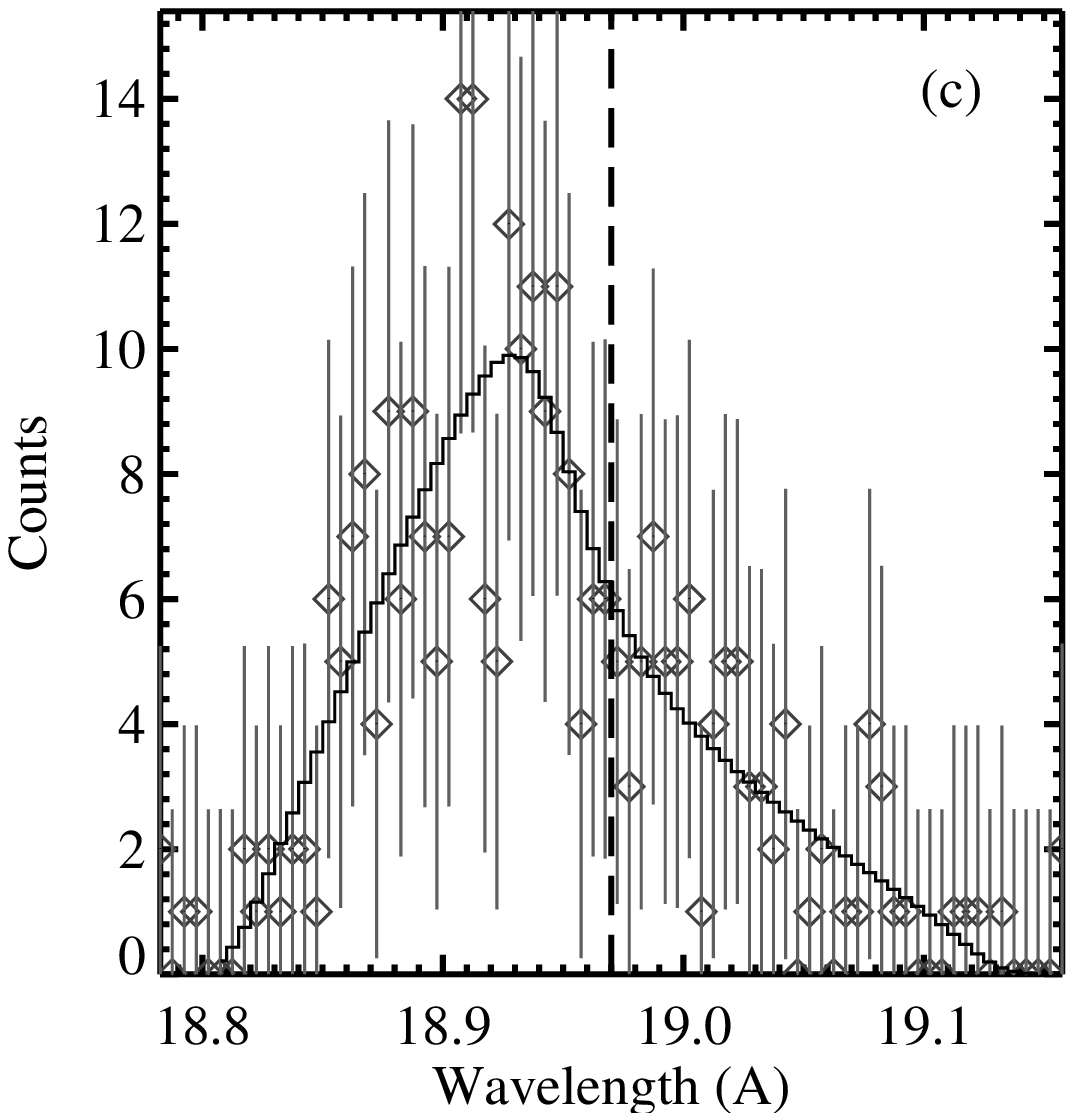}
\includegraphics[scale=0.3]{f1d.eps}
\caption{Two representative lines with best-fit models. Shown are
\chandra\ \meg\ spectra of (a) \ion{Fe}{17} at 15.01~\AA\ and (c)
\ion{O}{8} at 18.97~\AA\ (in gray), with corresponding best-fit
profiles (in black). Laboratory rest wavelengths are indicated by the
vertical dashed lines. To the right we show (b and d) contour plots
representing the models. The observer is located at $(p/\Rstar = 0,
z/\Rstar = \infty)$. The inner circle is of radius $\Rstar$, the outer
circle of radius $\Ro$. Gray contours are curves of constant
line-of-sight velocity component in units of $\vinf$. Dashed contours
are curves of constant optical depth ($\tau = 0.5, 1, 2$), integrated
along the line of sight.}
\label{fig:spectra}
\end{figure}

\begin{figure}
\includegraphics[scale=0.35]{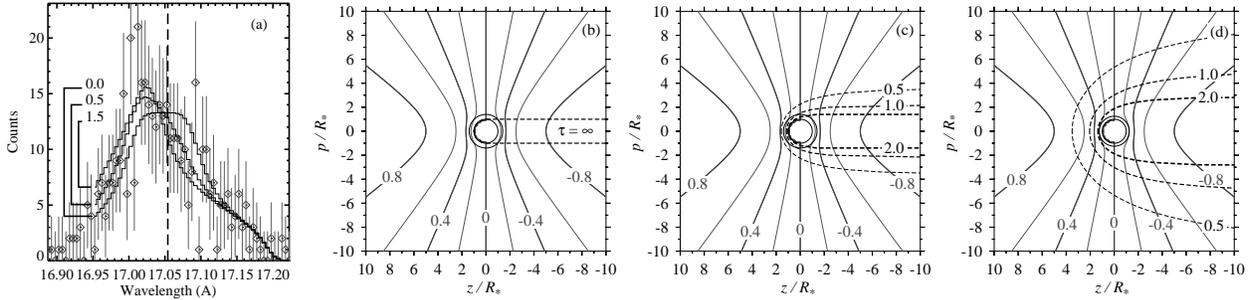}
\includegraphics[scale=0.3]{f2b.eps}
\includegraphics[scale=0.3]{f2c.eps}
\includegraphics[scale=0.3]{f2d.eps}
\caption{Models at extremes of the confidence region. Shown are the
\chandra\ \meg\ spectrum of the 17.054~\AA\ line of \ion{Fe}{17} (in
gray), and (in black) the best-fit model ($\taustar = 0.5$), the fit
with $\taustar$ held at its 95.4\% confidence upper limit ($\taustar =
1.5$) and the fit with $\taustar$ held at its 95.4\% confidence lower
limit ($\taustar = 0.0$). The contour plots are the same style as in
Figure \ref{fig:spectra} and correspond to (b) the $\taustar = 0.0$
model, (c) the best-fit model, and (d) the $\taustar = 1.5$ model.}
\label{fig:threemods}
\end{figure}

This model assumes implicitly that the sites of X-ray emission are so
numerous and well-mixed with the primary cool wind component that we
can treat the wind as a two-component fluid. It also neglects
non-radial velocity components, including small scale fluctuations
like turbulence. Note that the assumption of purely radial velocities
is what allows wind absorption to break the symmetry of the line,
since redshifted emission always arises at higher line-of-sight
distances (and therefore higher optical depths) than the blueshifted
emission (see the contour plots in Figures \ref{fig:spectra} and
\ref{fig:threemods}). 

Similar treatments of radiation transport have previously been fitted
to the global form of low-resolution spectra
\citep{hillier93,feldmeier97}. \citet{m91} and \citet{wc01}
have modeled line profiles from discrete spherical shocks. Elsewhere
we report on the investigation of non-spherical models and their
applicability to hot stars \citep{k03,t03}.

For this study, we have adopted the \citet{oc01} model and performed
fits on eight strong lines in the \chandra\ \hetgs\ spectrum of the O4
supergiant \zpup, extracting best-fit values and associated confidence
limits for three model parameters: $q$, $\Ro$, and $\taustar$.

\section{Fitting the Model to Observed Line Profiles}

Our data set consists of the $\pm 1$ order \meg\ spectrum from a $67$
ks observation of the O4f star \zpup\ first reported on by
\citet{cass01}. The FWHM of the \meg\ spectral response is $\Delta
\lambda_\mathrm{MEG} = 0.023$~\AA\ \citep{chandraguide}\footnote{The
\textit{Chandra} Proposers' Observatory Guide is available at
\url{http://cxc.harvard.edu/udocs/docs/}}. All the distinguishable
lines in this spectrum are many times more broad, allowing their
profiles to be well resolved. The breadth of the lines means that many
of them are contaminated by emission from neighboring lines. After
eliminating He-like \textit{fir} triplets as unsuitable for fitting due to
excessive blending, we identified other potential blends by visual
inspection of the spectrum and by referring to the line strengths
calculated by \citet[Table IV]{mewe85} and those in the Astrophysical
Plasma Emission Database \citep[APED,\ ][]{aped}\footnote{The
Interactive GUIDE for ATOMDB is available at
\url{http://obsvis.harvard.edu/WebGUIDE/}}. We consider a line with
rest wavelength $\lambda_0$ to extend over a wavelength range defined
by
\begin{equation}
\lambda_0 \left(1+\frac{v_\infty}{c} \right)+\Delta \lambda_\mathrm{MEG} \geq
\lambda \geq \lambda_0 \left(1-\frac{v_\infty}{c}\right)-\Delta
\lambda_\mathrm{MEG}
\;\mathrm{.}
\end{equation}
The widths of neighboring lines are calculated the same way, and any
overlap in the ranges is excluded from the fit (see Table
\ref{results_table} for the wavelength range over which each fit was
performed). We adopted the terminal velocity value determined by
\citet{prinja90}, $\vinf = 2485\;\mathrm{km\ s}^{-1}$.

We numerically integrate equation (\ref{L_eq}) over the desired
wavelength range using software written in \textit{Mathematica}. The
resulting profile is convolved with a Gaussian representing
instrumental response, binned identically to the data, and normalized
to predict the same total number of counts as were observed over the
same wavelength range.

To properly treat the statistics of Poisson-distributed,
low-photon-count data, we use Cash's $C$ \citep{c79} as the fit
statistic. Fits are performed by calculating $C$ on a grid in
parameter space. The coordinates of the parameter-space point
producing the minimum $C$ value $C_\mathrm{min}$ are taken as our
``best-fit'' parameters. We then use limits on $\Delta C = C -
C_\mathrm{min}$ to define our confidence regions, as described by
\citet{c79}. The grid is expanded as needed until the entire
confidence region is encompassed within it. In all our fits we held
$\beta$ constant at $\beta=1$ \citep[which is very close to the
typical O-star value of 0.8, ][]{groen89} and varied $\Ro$, $q$, and
$\taustar$. 

To confirm the confidence limits derived using the $\Delta C$
statistic, we carried out the fitting procedure on monte carlo
simulated data sets and compared the parameter-space distribution of
simulated-data fit parameters to the calculated $\Delta C$ confidence
region for each line. The monte carlo simulations gave results that
were consistent with those given by the $\Delta C$ statistic. 

The quality of the fits was evaluated using the Kuiper statistic, a
variant of the Kolmogorov-Smirnov test. Significance levels were
determined using monte carlo techniques. We obtained two slightly
different distributions of the Kuiper statistic depending on whether
the simulated data sets were compared directly to the parent model
(giving significance level $\alpha_0$) or compared to a new best-fit
model found by performing the fitting procedure on the simulated data
set (giving significance level $\alpha_\mathrm{RF}$).  

In all, we fit eight lines between 6.18~\AA\ (\ion{Si}{14}) and
24.78~\AA\ (\ion{N}{7}). The results are listed in Table
\ref{results_table}, and shown in Figure \ref{fig:spectra} for two
representative lines. All the fits but one are formally good according
to both distributions. The value of $\alpha_0 = 0.039$ for the
15.262~\AA\ fit does not meet the criterion $\alpha \ge 0.05$ for a
formally-good fit. This line may be contaminated by the \ion{Fe}{19}
lines at 15.198~\AA\ and 15.3654~\AA\ (APED). The relatively-low
significance values for the \ion{N}{7} line may simply be the result
of random variation, or could be a sign that there are resolved
spectral features not explained by this simple model. In any case, the
fit is formally good. The 16.787~\AA\ fit may also be affected by a
blend with an \ion{Fe}{19} line, this one at 16.718~\AA. This would
add flux to the blue edge of the line, increasing its skew and
explaining the relatively high values of $\taustar$ and $q$.

To demonstrate the typical range of models that can be fit to one
line, we show the best-fit and two extreme models superimposed on the
\ion{Fe}{17} line at 17.05~\AA\ in Figure \ref{fig:threemods}. The two
extreme models are for the parameter sets that have the largest and
smallest values of $\taustar$ within the 95.4\% confidence region.

\begin{figure}
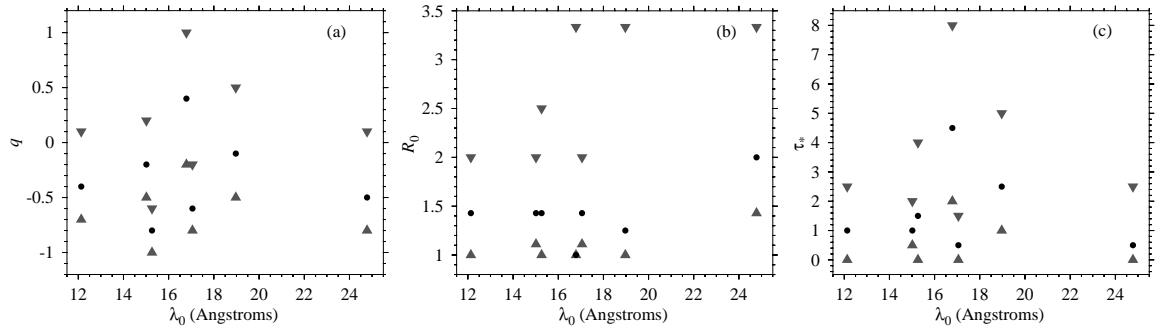

\includegraphics[scale=0.33]{f3a.eps}
\includegraphics[scale=0.33]{f3b.eps}
\includegraphics[scale=0.33]{f3c.eps}
\caption{The best-fit values of (a) $q$, (b) $R_o$ in units of $R_*$,
and (c) $\taustar$ for seven of the eight lines (bullets), along with
the range given by the 95.4\% confidence limits (triangles). See Table
\ref{results_table}. \label{fig:fits}}
\end{figure}

Finally, we summarize the best-fit and 95.4\% confidence limits of the
three free model parameters for seven of the eight lines in Figure
\ref{fig:fits} (the values for the \ion{Si}{14} line are not shown
because it is at a much shorter wavelength and its fit parameters are
poorly constrained).  Trends in these fitted parameters and their
implications are discussed in the next section.

\section{Discussion}

The primary result of the analysis presented in this paper is that the
X-ray emission lines in the prototypical O supergiant, \zpup, can, for
the most part, be adequately fit with a spherically symmetric wind
model having a small number of free parameters. Furthermore, the derived
parameters are quite reasonable in the context of most wind-shock
models, being consistent with hot plasma uniformly distributed
throughout the wind above a moderate onset radius, X-ray emitting
plasma extending out to the wind terminal velocity, and the need for
the inclusion of some wind attenuation.

In detail, however, some interesting trends emerge. First of all, the
amount of wind attenuation is significantly smaller than what one
might expect from a spherically symmetric smooth wind, given what is
known about this star's mass-loss rate and wind opacity. There have
been various calculations of the wind optical depth (often expressed
as the radius of optical depth unity) as a function of wavelength for
this star. They range from values much bigger than what we derive here
\citep[$7 < \taustar < 30$ calculated by][using $\mdot = 5.0 \times
10^{-6}~\msunyr$, $R_* = 19~R_\sun$, $\vinf =
2200$~km~s$^{-1}$]{hillier93}, to values modestly bigger \citep[$4 <
\taustar < 8$ calculated by][using values from \citealp{lamers93}:
$\mdot = 2.4 \times 10^{-6}~\msunyr$, $R_* = 16~R_\sun$, $\vinf =
2200$~km~s$^{-1}$]{cass01}. Note that \citet{hillier93} find different
values for $R_1$ depending on whether helium recombines or remains
ionized in the outer wind, but at energies above 0.5~keV (where all of
the lines presented here occur) there is little difference between the
two scenarios. More recent stellar parameters determined by
\citep{puls1996} ($\mdot = 5.9 \times 10^{-6}~\msunyr$, $R_* =
19~R_\sun$, and $\vinf = 2250$~km~s$^{-1}$) agree well with the values
used by \citet{hillier93}, but would increase the \citet{cass01}
$\taustar$ values by a factor of 2, given the same opacity (see
eq. \ref{taustar_eq}).

If we accept the $\taustar$ values derived from our fits, then the
disparity between those values and the ones mentioned above suggest
that either the mass-loss rates or wind opacities are being
overestimated in previous calculations. The mass-loss rate of \zpup\
is by now quite well established using UV absorption lines and
H$\alpha$, although improper ionization corrections or clumping could
lead to systematic errors. The wind opacity determination seems much
more uncertain, both because of the inconsistent values in the
literature and because of the difficulty in determining the ionization
state of the wind \citep{mac93,mac94}. Recent advances in stellar
atmosphere modeling may help to improve these determinations
\citep{pauldrach2001}. Another means of lowering the wind attenuation
is to clump the wind into small clouds that are individually optically
thick rendering the wind porous and enhancing the escape probability
of X-ray photons, thus lowering the mean wind opacity. This would also
affect the mass-loss rate diagnostics, but is, itself, an independent
effect.

An even more curious result of the $\taustar$ fits is that they are
nearly independent of wavelength. This is surprising because
photoionization cross sections should scale roughly as a power of
wavelength between $\lambda^2$ and $\lambda^3$ \citep{hillier93}.  It
is possible that the distribution of ionization edges could conspire
to make this relationship much flatter over a small range of
wavelengths (as the calculations from \citet{cass01} seem to
indicate). But wind clumping might play some role, here too. If the
wind opacity is dominated by clumps that are individually optically
thick across the wavelength range, then the opacity ceases to be a
function of wavelength and instead depends on the physical cross
sections of the clumps themselves. We note that the UV line opacity
necessary to explain the observed absorption line profiles could, in
principle, still be provided by the tenuous inter-clump wind, as the
line cross sections are much bigger than the X-ray photoionization
cross sections.

If $R_1 \gg \Ro$, the line profile is insensitive to changes in $\Ro$,
since emission much below $R_1$ is largely absorbed by the wind. The
values we find for $\Ro$, though generally small, are comparable to our
values of $R_1$ (from $\taustar$ by eq. \ref{r1_eq}). It is hard to
assess these relatively small onset radii in the context of the small
(sometimes surprisingly small) values claimed on the basis of observed
$f/i$ ratios in He-like ions \citep{cass01, k01, wc01}. This is
partially because we do not fit the profiles of any He-like lines
(they are too blended) and partly because the most extreme results
(smallest value for $R_1$) are for \ion{S}{15}, which is a higher
ionization stage than any of the lines we fit.

The fit results for the parameter $q$ indicate that there is not a
strong radial trend in the filling factor. One might expect some
competition in a wind shock model between the tendency to have more
and stronger shocks near the star, where the wind is still
accelerating, and the tendency for shock heated gas to cool less
efficiently in the far wind, where densities are low. Perhaps these
two effects cancel to give the observed $q \approx 0$ relationship.

In conclusion, the simple, spherically symmetric wind shock model is
remarkably consistent with the observed line profiles in \zpup,
providing the most direct evidence yet that some type of wind-shock
model applies to this hot star. However, there are indications that
the absorption properties of the wind of \zpup, and perhaps other hot
stars, must be reconsidered. We will fit this same model to the
\chandra\ spectra of other hot stars in the future. But the lack of
strong line asymmetries in stars such as $\zeta$ Ori and $\delta$ Ori
and the narrow lines in $\theta^1$ Ori C and $\tau$ Sco indicate that
spherically symmetric wind-shock models with absorption may not fit
the data from these stars as well as they do \zpup.

\acknowledgements

We wish to acknowledge grant GO0-1089A to Swarthmore College and Prism
Computational Sciences. RHK acknowledges funding provided by a Howard
Hughes Medical Institute undergraduate research grant. SPO
acknowledges partial support by NSF grant AST00-97983 and NASA grant
NAG5-3530 to the bartol Research Institute of the University of
Delaware. We would also like to thank the referee, Rolf-Peter
Kudritzki, for his comments which substantially improved the paper.

\begin{deluxetable}{lrrrrrrrrrrr}
  \tablewidth{0pt} \tablecolumns{12} 
\rotate
\tablecaption{\label{results_table}Best-Fit Parameters with 95.4\%
  Confidence Limits}
  \tablehead{
\colhead{Ion} &
\colhead{$\lambda_0$ (\AA)}&
\colhead{$\lambda_0/\Delta \lambda$}&
\colhead{$2\vinf/\Delta v$}&
\colhead{$q$}&
\colhead{$\Ro$}&
\colhead{$\taustar$}&
\colhead{$x_\mathrm{min}$}&
\colhead{$x_\mathrm{max}$}&
\colhead{$N_\mathrm{obs}$}&
\colhead{$\alpha_0$}&
\colhead{$\alpha_\mathrm{RF}$}
        }
  \startdata
\ion{N}{7}&	     $24.78$&	       $1077$&	                                      $18$&	                                                      $-0.5_{-0.3}^{+0.6}$&	   $2._{-0.6}^{+1.3}$&	    $0.5_{-0.5}^{+2.}$&	    $-1.11$&	            $1.11$&	             $92$&	$0.667$&	$0.356$\\
\ion{O}{8}&	     $18.97$&	       $825$&	                                       $14$&	                                                      $-0.1_{-0.4}^{+0.6}$&	   $1.2_{-0.2}^{+2.1}$&	   $2.5_{-1.5}^{+2.5}$&	   $-1.04$&	            $1.14$&	             $69$&	$0.941$& $0.790$\\
\ion{Fe}{17}&	   $17.054$&	      $741$&	                                       $12$&	                                                      $-0.6_{-0.2}^{+0.4}$&	   $1.4_{-0.3}^{+0.6}$&	   $0.5_{-0.5}^{+1.}$&	    $-0.74$&	            $1.16$&	             $54$&	   $0.971$&	      $0.950$\\
\ion{Fe}{17}\tablenotemark{a}&	   $16.787$&	      $730$&	                                       $12$&	                                                      $0.4_{-0.6}^{+0.6}$&	    $1.0_{-0}^{+2.3}$&	      $4.5_{-2.5}^{+3.5}$&	   $-1.16$&	            $0.74$&	             $53$&	   $0.517$&	      $0.490$\\
\ion{Fe}{17}\tablenotemark{a}&	   $15.262$&	      $664$&	                                       $11$&	                                                      $-0.8_{-0.2}^{+0.2}$&	   $1.4_{-0.4}^{+1.1}$&	   $1.5_{-1.5}^{+2.5}$&	   $-0.81$&	            $1.19$&	             $50$&	   $0.039$&	      $0.151$\\
\ion{Fe}{17}&	   $15.013$&	      $653$&	                                       $11$&	                                                      $-0.2_{-0.3}^{+0.4}$&	   $1.4_{-0.3}^{+0.6}$&	   $1.0_{-0.5}^{+1.}$&	     $-1.19$&	            $0.79$&	             $49$&	   $0.887$&	      $0.800$\\
\ion{Ne}{10}&	   $12.13$&	       $527$&  $9$&  $-0.4_{-0.3}^{+0.5}$&	   $1.4_{-0.4}^{+0.6}$&	   $1.0_{-1.}^{+1.5}$&  $-1.23$&	            $0.03$&	             $26$&	$0.432$&	      $0.100$\\
\ion{Si}{14}\tablenotemark{b}&	   $6.18$&	        $269$&	                                       $4$&	                                                       $-0.2_{-0.8}^{+\;\cdots}$&	   $1.4_{-1.4}^{+8.6}$&	   $1.5_{-1.5}^{+5.5}$&	   $-0.98$&	            $0.59$&	             $16$&	    $0.857$&	      $0.622$\\
  \enddata
\tablecomments{
The width of the instrumental response in wavelength units is $\Delta
\lambda = \Delta \lambda_\mathrm{MEG} = 0.023$~\AA, or in
velocity units $\Delta v = c \Delta \lambda / \lambda_0$. The scaled
wavelength $x \equiv  (c/\vinf) (\lambda-\lambda_0) /
\lambda_0$. $N_\mathrm{obs}$ is the number of wavelength bins included
in the fit. 
}
\tablenotetext{a}{The anomalous results for these lines may be due to
  contamination.}
\tablenotetext{b}{At the 95.4\% confidence level upper limit, $q$ is
  unconstrained for this fit.}
\end{deluxetable}

\newpage

\begin{thebibliography}{}

\bibitem[Babel \& Montmerle(1997)]{bm97} Babel, J., \& Montmerle, T. 1997, \apjl, 485, L29

\bibitem[Cash(1979)]{c79} Cash, W. 1979, \apj, 228, 939

\bibitem[Cassinelli \etal{}(2001)]{cass01} Cassinelli, J. P., Miller,
  N. A., Waldron, W. L., MacFarlane, J. J., \& Cohen, D. H. 2001,
  \apj, 554, L55

\bibitem[Cassinelli \& Olson(1979)]{co79} Cassinelli, J. P., \& Olson,
  G. L. 1979, \apj, 229, 304

\bibitem[\chandra\ X-Ray Center (2001)]{chandraguide} \chandra\ X-Ray Center.
2001, The \textit{Chandra} Proposers' Observatory Guide, v4.0,
(Cambridge: \chandra\ X-Ray Center)

\bibitem[Chen \& White(1991)]{cw91} Chen, W., \& White, R. L. 1991,
  \apj, 366, 512
  
\bibitem[Cohen \etal{}{}(1996)]{cohen96} Cohen, D. H., Cooper, R. G.,
  MacFarlane, J. J, Owocki, S. P., Cassinelli, J. P., \& Wang, P. 1996,
  \apj, 460, 506
  
\bibitem[Cohen \etal{}(2003)]{cohen03} Cohen, D. H., De M\`essieres, G.
  E., MacFarlane, J. J, Miller, N. A., Cassinelli, J. P., Owocki, S.
  P., \& Liedahl, D. A. 2003, \apj, in press

\bibitem[Feldmeier \etal{}(1997)]{feldmeier97} Feldmeier, A.,
Kudritzki, R.-P., Palsa, R., Pauldrach, A. W. A., Puls, J. 1997, \aap,
320, 899

\bibitem[Feldmeier, Puls, \& Pauldrach(1997)]{fpp97} Feldmeier, A.,
  Puls, J., \& Pauldrach, A. W. A. 1997, \aap, 322, 878

\bibitem[Gagn\'e \etal{}(1997)]{g97} Gagn\'e, M., Caillault, J.-P.,
  Stauffer, J. R., \& Linsky, J. L. 1997, \apj, 478, L87

\bibitem[Groenewegen \etal{}(1989)]{groen89} Groenewegen, M. A. T.,
Lamers, H. J. G. L. M., \& Pauldrach,
A. W. A. 1989, \aap, 221, 78

\bibitem[Harnden \etal{}(1979)]{h79} Harnden, F. R., Jr. \etal{}\ 1979,
  \apj, 234, L51

\bibitem[Hillier \etal{}(1993)]{hillier93} Hillier, D. J., Kudritzki
R. P., Pauldrach, A. W., Baade, D., Cassinelli, J. P., Puls, J., \&
Schmitt, J. H. M. M. 1993, \aap, 276, 117

\bibitem[Ignace(2001)]{i01} Ignace, R. 2001, \apjl, 549, L119

\bibitem[Kahn \etal{}(2001)]{k01} Kahn, S. M., Leutenegger, M. A.,
  Cotam, J., Rauw, G., Vreux, J.-M., den Boggende, A. J. F., Mewe, R.,
  \& G\"udel, M. 2001, \aap, 365, L312

\bibitem[Kramer \etal{}(2003)]{k03} Kramer, R. H., Tonnesen, S. K.,
  Cohen, D. H., Owocki, S. P., ud-Doula, A., \& MacFarlane,
  J. J. 2003, {\it Rev. Sci. Inst.}, in press

\bibitem[Lamers \& Leitherer(1993)]{lamers93} Lamers, H. J. G. L. M., \&
Leitherer, C. 1993, \apj, 412, 771

\bibitem[MacFarlane \& Cassinelli(1989)]{mc89} MacFarlane, J. J.,
  \& Cassinelli, J. P. 1989, \apj, 347, 1090

\bibitem[MacFarlane \etal{}(1991)]{m91} MacFarlane, J. J., Cassinelli,
  J. P., Welsh, B. Y., Vedder, P. W., Vallerga, J. V., \& Waldron, W. L. 1991, \apj, 380, 564

\bibitem[MacFarlane \etal{}(1993)]{mac93} MacFarlane, J. J., Waldron,
W. L., Corcoran, M. F., Wolff, M. J., Wang, P., \& Cassinelli,
J. P. 1993, \apj, 419, 813

\bibitem[MacFarlane, Cohen, \& Wang(1994)]{mac94} MacFarlane, J. J.,
Cohen, D. H., \& Wang, P. 1994, \apj, 437, 351

\bibitem[Mewe, Gronenschild, \& van den Oord(1985)]{mewe85} Mewe, R.,
Gronenschild, E. H. B. M., \& van den Oord, G. H. J. 1985, \aaps, 62, 197

\bibitem[Mighell(1999)]{m99} Mighell, K. J. 1999, \apj, 518, 380

\bibitem[Miller \etal{}(2002)]{m03} Miller, N. A., Cassinelli, J. P.,
Waldron, W. L., MacFarlane, J. J., \& Cohen, D. H. 2002, \apj, 577, 951

\bibitem[Owocki, Castor, \& Rybicki(1988)]{ocr88} Owocki, S. P.,
  Castor, J. I. \& Rybicki, G. B. 1988, \apj, 335, 914
  
\bibitem[Owocki \& Cohen(2001)]{oc01} Owocki, S. P., \& Cohen, D. H.
  2001, \apj, 559, 1108

\bibitem[Pauldrach, Hoffmann, \& Lennon(2001)]{pauldrach2001}
Pauldrach, A.~W.~A., Hoffmann, T.~L. \& Lennon, M. 2001, \aap, 375,
161

\bibitem[Prinja, Barlow, \& Howarth(1990)]{prinja90} Prinja, R. K., Barlow,
M. J., \& Howarth, I. D. 1990, \apj, 361, 607

\bibitem[Puls \etal{}(1996)]{puls1996} Puls, J., Kudritzki, R.-P.,
Herrero, A., Pauldrach, A. W. A., Haser, S. M., Lennon, D. J., Gabler,
R. Voels, S. A., Vilchez, J. M., Wachter, S., Feldmeier, A. 1996,
\aap, 305, 171

\bibitem[Schulz \etal{}(2001)]{s01} Schulz, N. S., Canizares, C. R.,
  Huenemoerder, D. \& Lee, J. C. 2001, \apjl, 549, 441.
  
\bibitem[Smith \etal{}(2001)]{aped} Smith, R. K., Brickhouse, N. S., Liedahl, D. A., \& Raymond, J. C. 2001, \apjl, 556, L91
  
\bibitem[Tonnesen \etal{}(2003)]{t03} Tonnesen, S. K., Cohen, D. H.,
Owocki, S. P., ud-Doula, A., Gagne, M., \& Oksala, M. 2003, \baas, 34,
1284

\bibitem[ud-Doula \& Owocki(2002)]{uo02} ud-Doula, A., \& Owocki,
  S. P. 2002, \apj, 576, 413

\bibitem[Waldron(1984)]{w84} Waldron, W. L. 1984, \apj, 282, 256

\bibitem[Waldron \& Cassinelli(2001)]{wc01} Waldron, W. L., \&
Cassinelli, J. P. 2001, \apjl, 548, L45
\end{thebibliography}
\end{document}